\documentclass{aa}
\usepackage{graphicx}
\usepackage{txfonts}
\usepackage{natbib}
\bibpunct{(}{)}{;}{a}{}{,} 

\begin{document}
\title{Observations of three pre-cataclysmic variables from the Edinburgh-Cape 
Blue Object Survey}
\titlerunning{Observations of three pre-cataclysmic variables}

\author{
C.~Tappert\inst{1}\thanks{new address: Departamento de F\'{\i}sica y 
Astronom\'{\i}a, Universidad de Valpara\'{\i}so, Avenida Gran Breta\~na 1111, 
Valpara\'{\i}so, Chile},
B.~T.~G\"ansicke\inst{2},
M.~Zorotovic\inst{1},
I.~Toledo\inst{1},
J.~Southworth\inst{2},
C.~Papadaki\inst{3},
R.~E.~Mennickent\inst{4}
}

\authorrunning{C. Tappert et al.}

\offprints{C. Tappert}

\institute{
Departamento de Astronom\'{\i}a y Astrof\'{\i}sica, Pontificia Universidad
Cat\'olica, Vicu\~na Mackenna 4860, 782-0436 Macul, Chile\\
\email{ctappert@dfa.uv.cl,mzorotov@astro.puc.cl,itoledoc@gmail.com}
\and
Department of Physics, University of Warwick, Coventry CV4 7AL, UK\\
\email{Boris.Gaensicke@warwick.ac.uk,j.k.taylor@warwick.ac.uk}
\and
Vrije Universiteit Brussel, Pleinlaan 2, 1050 Brussels, Belgium\\
\email{cpapadak@vub.ac.be}
\and
Departamento de Astronom\'{\i}a, Universidad de Concepci\'on, Casilla 160-C,
Concepci\'on, Chile\\
\email{rmennick@astro-udec.cl}
}

\date{Received xxx; accepted xxx}

\abstract
{} {This study aims at determining the parameters of the three
  candidate pre-cataclysmic binaries EC\,12477--1738, EC\,13349--3237, and
  EC\,14329--1625, most importantly their orbital period.}
{ 
Time-series photometry reveals orbital modulation in the form of
sinusoidal variation due to the reflection effect. Photometric
observations are complemented by time-resolved spectroscopy that
yields radial velocities of the H$\alpha$ emission line. The
combination of both methods allows us to unambiguously determine the
orbital periods. The average spectra are used to estimate physical
parameters of the primary and secondary stellar components.}
{
We determine the orbital period for EC\,12477--1738 as 0.362 d, thus confirming 
the value previously reported. A similar period, $P=0.350$\,d, is found for 
EC\,14329--1625. Both systems incorporate a medium-hot
white dwarf ($T = 15\,000-20\,000$ K) and an M3V secondary star. The third
pre-CV, EC\,13349--3237, is the youngest of the three, with a hot WD 
($T \sim 35\,000$ K), and it also has the longest period $P=0.469$ d. 
It furthermore turns out to be one of the still rare pre-CVs with\,a 
comparatively early-type, M1V, secondary star, which will eventually
evolve into a CV above the period gap. 
}
{}


\maketitle

\section{Introduction}

\begin{table}
\caption[]{General information on the three targets of this study.}
\label{info_tab}
\begin{tabular}{llll}
\hline\noalign{\smallskip}
object & R.A. (2000.0)$^1$ & DEC (2000.0)$^1$ & $V$ [mag]$^2$\\
\hline\noalign{\smallskip}
EC 12477--1738 & 12:50:22.1 & -17:54:47 & 16.2 \\
EC 13349--3237 & 13:37:50.8 & -32:52:23 & 16.3 \\
EC 14329--1625 & 14:35:45.7 & -16:38:17 & 14.9 \\
\hline\noalign{\smallskip}
\multicolumn{4}{l}{
1) from the SIMBAD database; 2) from \citet{kilkennyetal97-1}}
\end{tabular}
\end{table}

\begin{figure*}
\includegraphics[width=2.0\columnwidth]{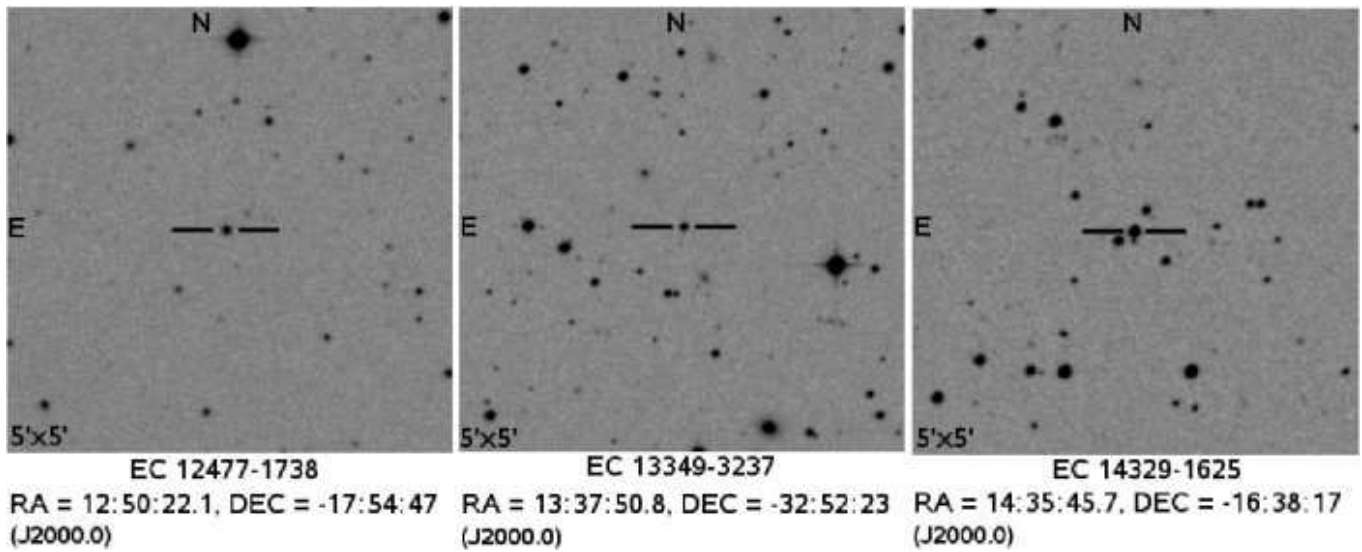}
\caption[]{Finding charts from the Digital Sky Survey (DSS2-red).}
\label{fcharts_fig}
\end{figure*}

Cataclysmic Variables (CVs) are close interacting binaries with a
white dwarf {accreting material from a
Roche-lobe filling, late-type (K--M), main-sequence star. These
systems are thought to form from initially separated binaries that go
through a common-envelope phase as the more massive 
one of the two stars expands in the course of its
nuclear evolution. The eventual expulsion of the
common-envelope material leaves a still separated
white dwarf / main-sequence star (hereafter WD/MS) binary.
Angular momentum loss due to magnetic
braking and gravitational radiation shrinks the separation, which
eventually leads to the secondary star filling its Roche lobe, thus
starting mass-transfer. Following \citet{schreiber+gaensicke03-1}, we
call the WD/MS binaries pre-CVs, if the evolution from the expulsion
of the common envelope to the start of mass transfer takes place
within Hubble time ($\sim$13 Gyr).

Recent attempts to solve the discrepancies between the modelled and the 
observed
CV population \citep[e.g., ][]{stehleetal97-1, patterson98-1, gaensicke05-1}
emphasise the importance of the pre-CV phase \citep{schenker+king02-1,
schenkeretal02-1}. The detection of anomalous element abundances
in a number of CVs \citep{gaensickeetal03-1,harrisonetal04-1,harrisonetal05-1}
raises the question of whether the secondary star undergoes a certain amount of 
nuclear evolution prior to entering the CV stage, although an
examination of 13 pre-CVs gave a negative result \citep{tappertetal07-2}.

Since pre-CVs consist of two intrinsically faint stellar components
and 
-- with the exception of eclipsing systems and binary central stars of
planetary nebulae -- 
show only very minor photometric variability (with amplitudes of
typically $\le$0.1 mag), they are not easily detected. Consequently
the number of known pre-CVs is small; applying the criteria from
\citet{schreiber+gaensicke03-1}, i.e., $P_{\rm orb} < 2~\mathrm{d}$
and a main-sequence secondary with $M_2 < M_1$, we find 56 potential
and confirmed pre-CVs in version 7.9 of the \citet{ritter+kolb03-1}
catalogue. The Sloan Digital Sky Survey (SDSS) is currently
fundamentally changing this picture, thanks to the vast number of
spectroscopic follow-ups of point sources with non-stellar $ugriz$
colours \citep{raymondetal03-1,silvestrietal06-1, silvestrietal07-1,
  rebassa-mansergasetal07-1, schreiberetal08-1}, and detailed
follow-up observations will eventually provide $\simeq2000$ new
systems. However, the majority of the SDSS systems will be faint,
requiring large-aperture telescopes for follow-up observations. 

Here, we present photometric and spectroscopic studies of three
relatively bright pre-CVs, \object{EC\,12477--1738}, \object{EC
  13349--3237}, and \object{EC\,14329--1625}, which were discovered and
classified as WD/MS binaries in the Edinburgh-Cape Blue Object Survey
\citep{kilkennyetal97-1}. 
Coordinates and apparent magnitudes of these
systems are collected in Table \ref{info_tab}; finding charts are presented
in Fig.\,\ref{fcharts_fig}.
The first attempts to derive their orbital
period photometrically have been described in \citet{tappertetal04-1}
and \citet{tappertetal06-2}.  \citet{maxtedetal07-2} used
time-resolved spectroscopy to determine the period for one of these
systems, EC\,12477--1738.

\section{Observations and data reduction}
\label{obs_sect}

\begin{table*}
\caption[]{Log of observations.}
\label{obs_tab}
\begin{tabular}{llllllll}
\hline\noalign{\smallskip}
object & date$^1$ & HJD$^1$ & telescope & configuration$^2$ 
& $n_\mathrm{data}$ & $t_\mathrm{exp}$ [s] & $\Delta t$ [h] \\
\hline\noalign{\smallskip}
EC\,12477--1738 & 2003-04-08 & 2\,452\,738 & 0.9 m & CCD+$R$ 
& 120 & 60/90    & 3.35 \\
              & 2003-04-09 & 2\,452\,739 & 0.9 m & CCD+$R$ 
& 70  & 90/180   & 4.08 \\
              & 2003-04-10 & 2\,452\,740 & 0.9 m & CCD+$R$ 
& 60  & 180      & 3.84 \\
              & 2007-04-03 & 2\,454\,194 & 4.0 m & KPGL3, 3565--7240, 4.8
& 7   & 900/1200 & 8.28 \\
              & 2007-04-05 & 2\,454\,196 & 4.0 m & KPGL3, 3565--7240, 4.8
& 9   & 900/1200 & 8.97 \\
EC\,13349--3237 & 2003-05-15 & 2\,452\,775 & 0.9 m & CCD+$R$
& 50  & 180/240  & 4.57 \\
              & 2003-05-16 & 2\,452\,776 & 0.9 m & CCD+$R$
& 47  & 240      & 3.68 \\
              & 2003-06-20 & 2\,452\,811 & 0.9 m & CCD+$R$
& 81  & 240      & 6.28 \\
              & 2003-06-21 & 2\,452\,812 & 0.9 m & CCD+$R$
& 77  & 240      & 5.98 \\
              & 2007-04-03 & 2\,454\,194 & 4.0 m & KPGL3, 3565--7240, 4.8
& 7   & 900/1200 & 7.59 \\
              & 2007-04-05 & 2\,454\,196 & 4.0 m & KPGL3, 3565--7240, 4.8
& 9   & 900/1200 & 8.67 \\
EC\,14329--1625 & 2003-06-18 & 2\,452\,809 & 0.9 m & CCD+$R$
& 119 & 90       & 4.16 \\
              & 2003-06-19 & 2\,452\,810 & 0.9 m & CCD+$R$
& 120 & 90       & 4.54 \\
              & 2005-04-25 & 2\,453\,486 & 0.9 m & CCD+$R$
& 9   & 150      & 3.57 \\
              & 2005-04-28 & 2\,453\,489 & 0.9 m & CCD+$R$
& 27  & 150      & 8.35 \\
              & 2005-04-29 & 2\,453\,490 & 0.9 m & CCD+$R$
& 36  & 150      & 7.88 \\
              & 2005-05-24 & 2\,453\,515 & 0.9 m & CCD+$R$
& 12  & 150/300  & 5.14 \\
              & 2005-05-26 & 2\,453\,517 & 0.9 m & CCD+$R$
& 14  & 180      & 4.95 \\
              & 2007-04-03 & 2\,454\,194 & 4.0 m & KPGL3, 3565--7240, 4.8
& 7   & 600/900  & 7.64 \\
              & 2007-04-05 & 2\,454\,196 & 4.0 m & KPGL3, 3565--7240, 4.8
& 8   & 600/900  & 8.37 \\
\hline\noalign{\smallskip}
\multicolumn{8}{l}{
1) start of night;
2) grism/grating, wavelength range, and spectral FWHM resolution in {\AA}
}
\end{tabular}
\end{table*}

The photometric data were taken on several occasions in 2003 and 2005
at the 0.9 m CTIO/SMARTS telescope using an $R$ filter. The 2003
observations were part of a survey on a sample of 16 objects that had been
classified as candidate pre-CVs based on their spectral appearance. The aim
was to examine them for potential photometric variability that would 
confirm their classification and reveal their orbital period
\citep{tappertetal04-1}. Since it was unknown if the targets would 
show any variation, continuous light curves were taken, i.e.\, a certain fraction 
of the night (usually around 4 h) would be dedicated to a specific
object. For the 2005 observations, the sample was limited to 5 objects with
known or suspected light curve modulations corresponding to periods $> 6$ h. 
Thus, targets were cycled throughout the night, with three consecutive data 
points being taken per step.  In this way, light curves covering $\sim$8 h per 
night could be measured for three to four  targets, yielding a time
resolution of $\sim$0.5 h. Unfortunately, the 2005 observing runs were
plagued with bad weather conditions, so that in the end only the data
for the brightest target, EC\,14329--1625, proved useful.

From the photometric observations, four systems emerged as confirmed
pre-CVs: \object{LTT 560} \citep{tappertetal07-1} and the three targets
of the present paper. The latter were selected for time-resolved spectroscopy
on 3 nights in April 2007 at the 4.0 m CTIO in order to pin down the
orbital period, since the photometry on its own does not allow for an unambiguous
distinction between sinusoidal (one maximum per orbit due to the secondary star
being irradiated by a hot white dwarf) or ellipsoidal variation (two maxima per orbit
due to the deformed secondary star). We employed the R-C spectrograph and grating
KPGL3 with a 1.0\arcsec slit to yield a wavelength range of 3565--7240 {\AA} 
at a spectral resolution of 4.8 {\AA}. 
Using essentially the same strategy as for the 2005 photometric observations,
after each target spectrum, a HeAr lamp wavelength calibration exposure was taken, 
and afterwards the telescope was pointed to the next object.
Flux calibration standards \object{LTT 4816} and
\object{LTT 7379} were observed at the beginning of night 1 and at the end
of night 3, respectively. Since the weather conditions were not
photometric (in fact, the middle night of our observing run was
completely overcast), one expects the respective calibrations not to
be very accurate.  However, as we will see below, they still provide
valuable information.  A summary of the observations is given in Table
\ref{obs_tab}.

Basic reduction of the photometric and spectroscopic data followed
standard procedures for bias subtraction and flat fielding, using
IRAF\footnote{NOAO PC-IRAF Revision 2.12.2-EXPORT} tasks. The majority
of the photometric data was analysed with IRAF's apphot/daophot
packages, and the stand-alone daomatch and daomaster routines
\citep{stetson92-1}. The final aperture radius for the photometry was
chosen as the one that gave a minimum noise in the differential light
curve of two non-variable field stars with magnitudes similar to the
target. All field star photometry of a single CCD frame were then
combined to give an average light curve, with variable and noisy stars
being iteratively excluded. The differential light curve for the
target finally was computed by subtracting the averaged field stars
from the target data. In an attempt to reduce the noise in the light
curves, an iterative weighting algorithm \citep{broegetal05-1} was
used to compute the average comparison light curve for the April 2005
data. However, the gain in signal-to-noise ratio (S/N) proved
insufficient to justify a re-evaluation of the other data sets.

The spectroscopic data were optimally extracted \citep{horne86-1} and
wavelength and flux calibrated. Radial velocities of the H$\alpha$
emission line (and of absorption lines of the secondary star, when
possible) were measured by fitting single Gaussians to the line
profile. The WD absorption lines were found to be too broad and the
S/N too low in order to be measured either by fitting or by using
cross-correlation methods. Additionally, narrow emission lines were
present in the centre of the Balmer absorption in all cases. The
H$\alpha$ radial velocities were corrected for the motion with
respect to the local standard of rest (LSR) as defined in IRAF's 
rvcorrect task (the Sun's velocity vector relative to the LSR being
$v_\odot = 20~\mathrm{km/s}$, $\alpha_{1900} = 18~\mathrm h$, 
$\delta_{1900} = 30^\circ$) and a potential instrumental flexure was 
accounted for by subtracting the variation of the 
[O\,{\sc I}] $\lambda$5577
night sky emission line.

Both the photometric and the spectroscopic data were examined for
periodic modulation applying several routines implemented in the
ESO-MIDAS\footnote{version 07FEBpl1.1 on PC/Linux} time-series
analysis (tsa) context, namely the \citet{scargle82-1} and
analysis-of-variance \citep[AOV; ][]{schwarzenberg-czerny89-1}
algorithms, as well as the extension of the latter, which uses
orthogonal trigonometric polynomials to fit the phase-folded data
\citep[ORT; ][]{schwarzenberg-czerny96-1}. After discarding alias
periods (see the respective sections for each system), the error 
associated with the selected period was calculated using equation 
(4) from \citet{larsson96-1}.

\section{Results}

\subsection{General spectroscopic appearance}
\label{genspec_sect}

\begin{figure*}
\includegraphics[angle=-90,width=2.0\columnwidth]{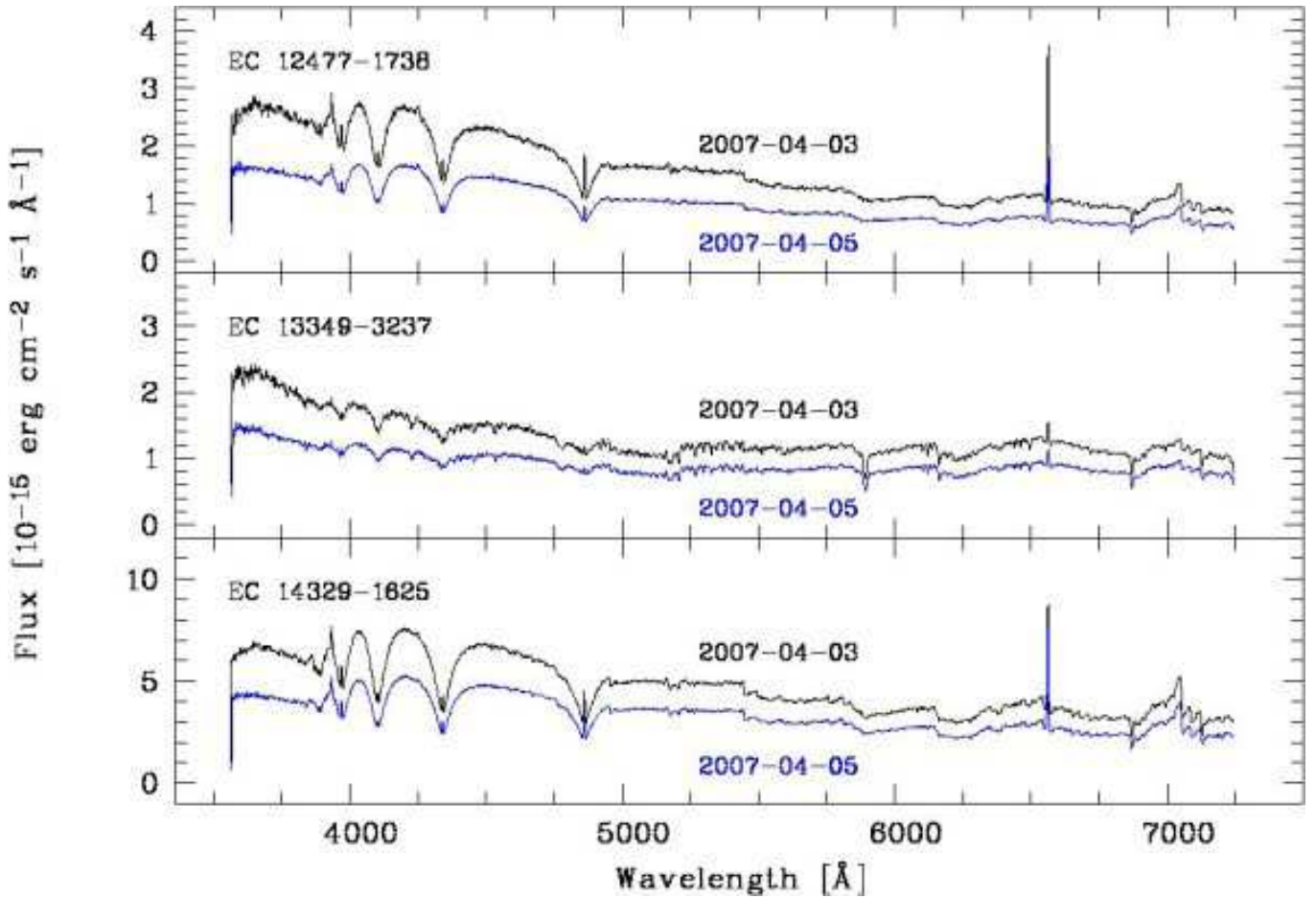}
\caption[]{Average spectra from April 3 and April 5, 2007 as labelled.
Note the composite nature of all three systems, with the white dwarf
dominating the blue part of the spectrum, and the late-type secondary
star showing its presence in the red part. The narrow hydrogen emission
lines also originate in the secondary star. The brightness and colour
differences between the two nights are not intrinsic. See text for more
detail.
}
\label{avsp_fig}
\end{figure*}

The nightly average spectra for all systems are presented in Fig.\
\ref{avsp_fig}. While the data for the three targets will be analysed
individually in the subsequent sections, here we
comment on the common spectroscopic properties. In all three objects,
the WD is dominant in the blue part of the spectrum.
EC\,12477--1738 and EC\,14329--1625 present broad and prominent Balmer 
absorption lines, while in EC\,13349--3237 they are not as pronounced due 
to the system incorporating a much hotter WD (see the respective section).

In the red part, the late-type secondary star shows its presence in the
form of a red continuum and molecular absorption bands. Again, while
EC\,12477--1738 and EC\,14329--1625 at first glance present a very similar
appearance, in EC\,13349--3237 the less pronounced TiO band edge around 
7050\,{\AA} and the presence of absorption features in the middle part of
the spectrum suggest an earlier spectral type for the secondary star than
in the other two systems.

We also find comparatively large differences for all spectra between one night
and the other. Specifically, the April 3 data are about 0.4 mag brighter
than the ones from April 5, and the continuum slopes appear redder on the
second night. We have already commented in Section \ref{obs_sect} on the
non-photometric conditions during the observations, and these therefore can
account for the brightness offset between the two nights. The
difference in colour is probably due to atmospheric refraction, since the
standard stars were observed at significantly different airmasses: 
LTT 4816 on April 3 at $M(z) \sim 1.7$ and LTT 7379 on April 5 close to the
zenith. In contrast, the targets have been observed throughout the night,
covering a wide range of airmasses, roughly from 2.0 to 1.0 for all three
systems. We therefore conclude that the observed differences both in
brightness and in colour are not intrinsic, but instead due to non-photometric
conditions and atmospheric refraction. However, as we will see in the 
next section, EC\,12477--1738 represents a special case.

\subsection{EC\,12477--1738}
\label{ec12477_sect}

\begin{figure}
\includegraphics[angle=-90,width=\columnwidth]{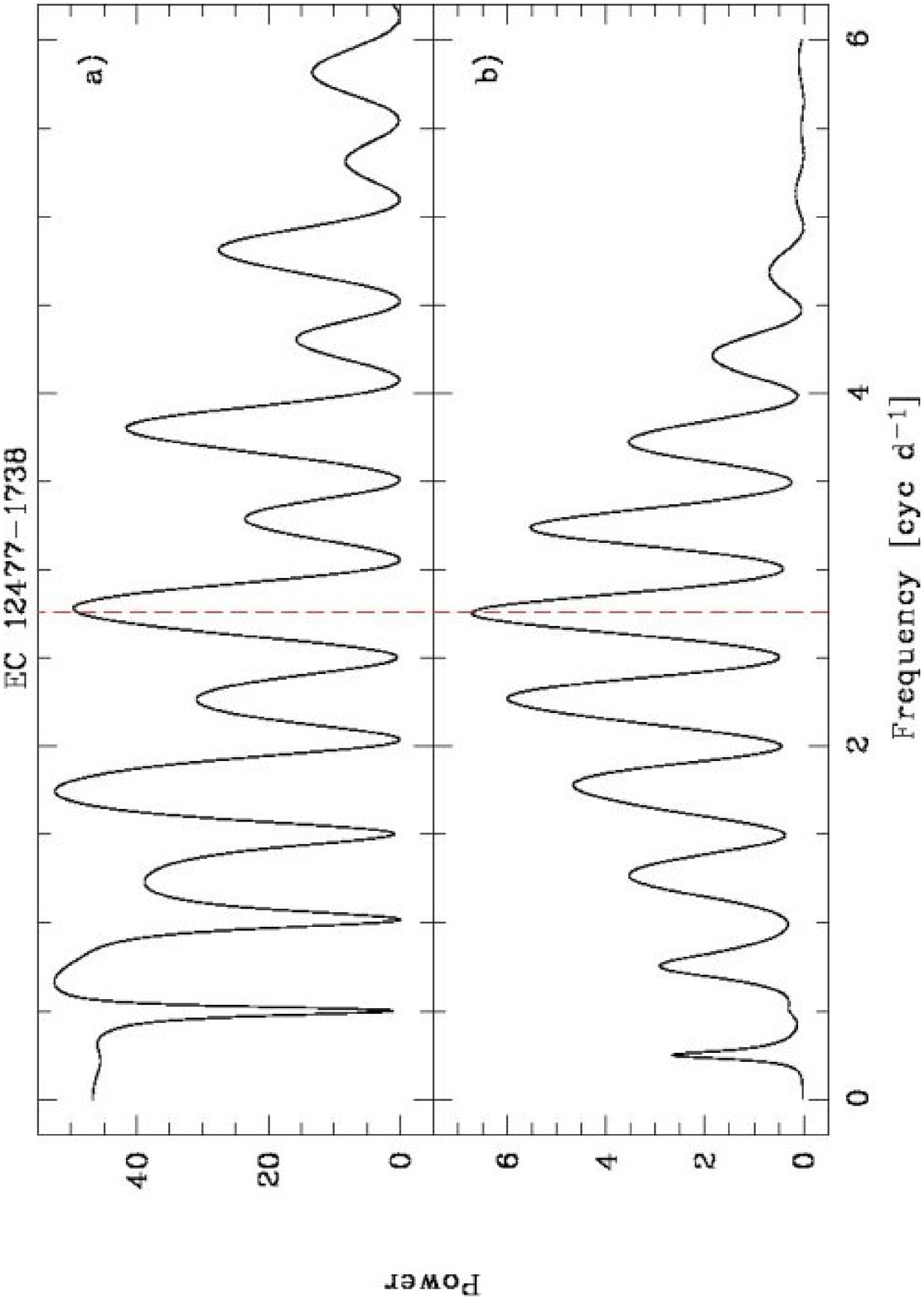}
\caption[]{Scargle periodograms for EC\,12477--1738. a) Photometric data. 
b) Radial velocities of the H$\alpha$ emission line.
The hashed line marks the adopted period 
$P = 0.362~\mathrm{d}$.}
\label{ec12pg_fig}
\end{figure}

\begin{figure}
\includegraphics[angle=-90,width=\columnwidth]{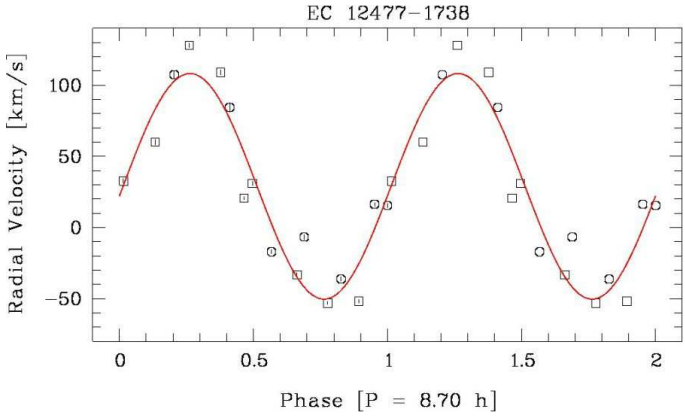}
\caption[]{Radial velocities 
of the H$\alpha$ emission line
for EC\,12477--1738, phase-folded with respect to
the ephemeris in Eq.\ \ref{ec12eph_eq}. Circles mark
data from 2007-04-03, squares those from 2007-04-05. The solid curve gives
the best sine fit. The first cycle includes error bars corresponding to the
Gaussian fit to the H$\alpha$ emission line.}
\label{ec12phvel_fig}
\end{figure}

\begin{figure}
\includegraphics[angle=-90,width=\columnwidth]{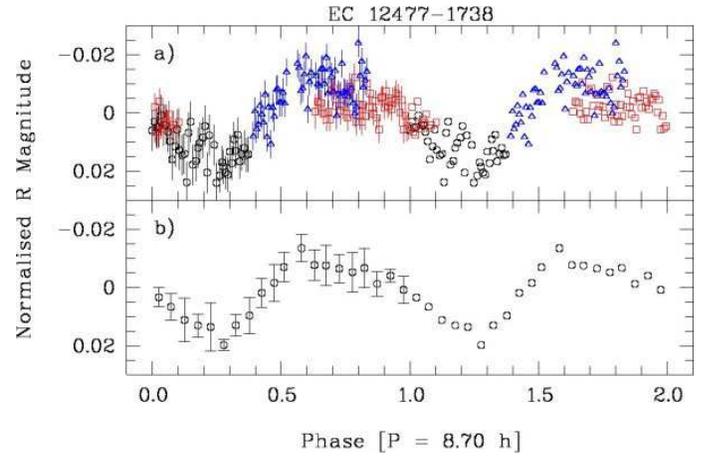}
\caption[]{Phase-folded light curves for EC\,12477--1738, with phase zero arbitrarily 
set to the first data point. a) The individual data
from April 08 (circles), 09 (squares), and 10 (triangles). Two cycles are shown,
the second one without error bars. b) The data averaged into bins of 0.05 phases. 
The error bar gives the standard deviation with respect to the average value.}
\label{ec12phlc_fig}
\end{figure}

\begin{figure}
\includegraphics[angle=-90,width=\columnwidth]{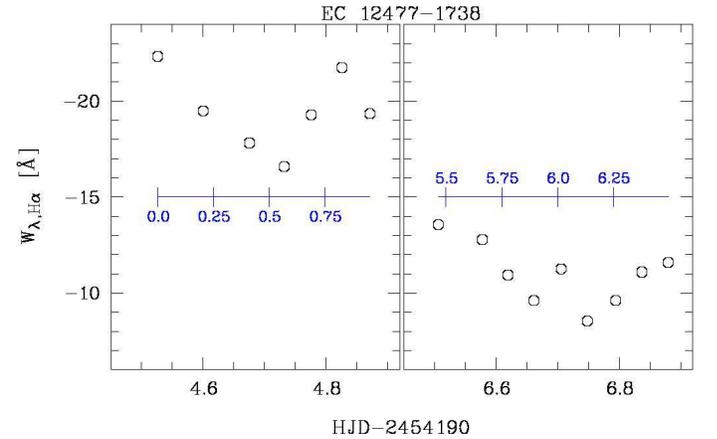}
\caption[]{H$\alpha$ equivalent widths of the individual EC\,12477--1738 
spectra. The tickmarks give the phase with respect to $P = 8.70$ h (first
data point = phase 0.0).}
\label{ec12eqw_fig}
\end{figure}

\begin{figure}[t]
\includegraphics[width=\columnwidth]{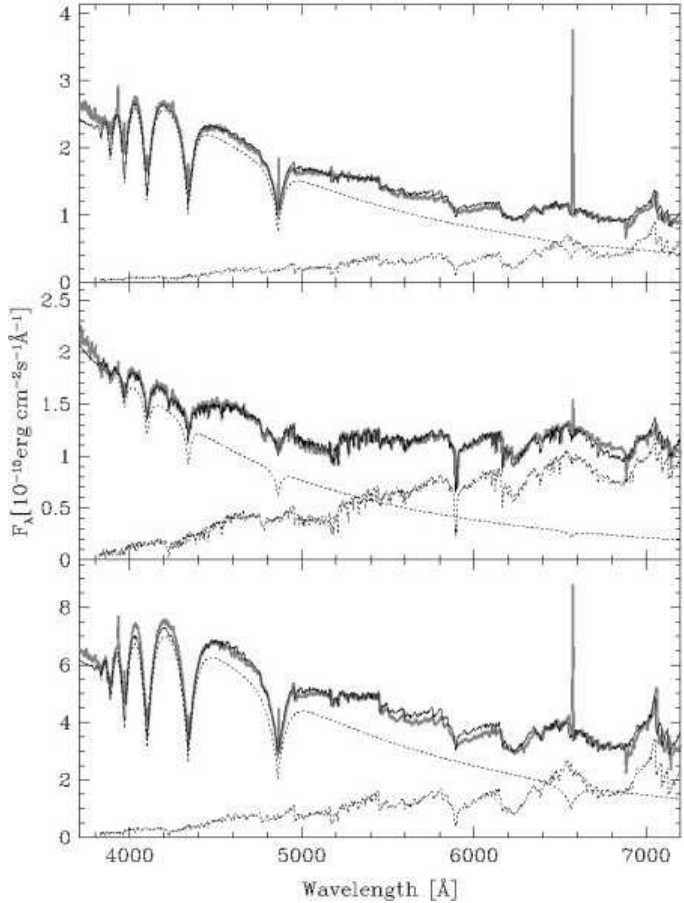}
\caption[]{Spectral fits to the spectra of EC\,12477--1738 (top
  panel), EC\,13349--3237 (middle panel) and EC\,14329--1625 (bottom
  panel). The observed data are plotted as a thick gray line, the
  best-fit M-dwarf as a dotted line, and the best-fit white dwarf as hashed
  line. The sum of both components is overplotted on the data as a thin
  black line.}
\label{ecspecfit_fig}
\end{figure}

\begin{table*}
\caption[]{Stellar parameters for the white dwarfs and M-dwarfs in
  EC\,12477--1738, EC\,13349--3237 and EC\,14329--1625 determined from
  a spectroscopic decomposition of the CTIO spectra. 
}
\label{specfit_tab}
\begin{tabular}{lcccccc}
\hline\noalign{\smallskip}
  & $T_\mathrm{wd}$\,[K] & $\log g$ & $M_\mathrm{wd}\,[M_\odot]$ 
  & $d_\mathrm{wd}$\,[pc] & Sp(sec) & $d_\mathrm{sec}$\,[pc] \\
\hline\noalign{\smallskip}
 EC\,12477--1738 & $18330\pm1212$ & $8.08\pm0.26$ & $0.67\pm0.16$ & $111\pm19$ & M$3\pm0.5$ & $206\pm41$\\
                 & $17106\pm1131$ & $7.88\pm0.26$ & $0.55\pm0.15$ & $151\pm24$ & M$3\pm0.5$ & $247\pm49$\\
 & $\it{17718\pm865}$ & $\it{7.98\pm0.14}$ & $\it{0.61\pm0.08}$ & $\it{131\pm28}$ & $\it{M3\pm0.5}$ & $\it{226\pm45}$ \\
 EC\,13349--3237 & $37424\pm3880$ & $7.73\pm0.54$ & $0.54\pm0.28$ & $394\pm133$& M$1\pm0.5$ & $321\pm63$\\
                 & $32595\pm2052$ & $7.31\pm0.53$ & $0.38\pm0.19$ & $572\pm196$& M$1\pm0.5$ & $373\pm73$\\
 & $\it{35010\pm3415}$ & $\it{7.52\pm0.30}$ & $\it{0.46\pm0.11}$ & $\it{483\pm126}$ & $\it{M1\pm0.5}$ & $\it{347\pm68}$ \\
 EC\,14329--1625 & $13904\pm1409$ & $7.87\pm0.25$ & $0.54\pm0.15$ & $71\pm11$  & M$3\pm0.5$ & $120\pm24$\\
                 & $15246\pm1299$ & $8.14\pm0.22$ & $0.70\pm0.14$ & $53\pm8$   & M$3\pm0.5$ & $106\pm21$\\
 & $\it{14575\pm949}$ & $\it{8.01\pm0.19}$ & $\it{0.62\pm0.11}$ & $\it{62\pm13}$ & $\it{M3\pm0.5}$ & $\it{113\pm10}$ \\
\hline\noalign{\smallskip}
\multicolumn{7}{l}{
\parbox{13cm}{
Note: Three lines are given for each object, with the top (middle) line 
referring to the parameters determined from the spectra taken on April 3 (5)
2007. The bottom line, in italics, gives the average values and sample 
variance, except for $d_\mathrm{sec}$, where the error is the average 
of the two individual errors.
}}
\end{tabular}
\end{table*}

The photometric data present clear variability, but have insufficient
coverage to pin down the orbital period, as is evident from the
corresponding periodogram (Fig.\ \ref{ec12pg_fig}a). The radial
velocities
of the H$\alpha$ emission line,
instead, yield a clear main peak at $f = 2.76~\mathrm{cyc~
d^{-1}}$, which corresponds to $P = 0.362~\mathrm{d} =
8.70~\mathrm{h}$ (Fig.\ \ref{ec12pg_fig}b). The spectroscopic periodogram
shows a number of alias periods with peak values larger than half the values 
of the strongest signal, which were also tested on the spectroscopic 
and photometric data. For all alias periods, at least one of the two 
phase-folded data sets showed strong discrepancies, especially regarding 
data from different nights, thus leaving the aforementioned period
corresponding to the strongest peak as the only viable choice.

Figs.\ \ref{ec12phvel_fig} and \ref{ec12phlc_fig} give the
phase-folded radial velocities and the light curve, respectively. The
corresponding sine fit to the velocities $v(\varphi)$,
\begin{equation}
v(\varphi) = \gamma + K_2 \sin \varphi
\end{equation}
yields $\gamma = 29(03)~\mathrm{km~s^{-1}}$, $K_2 = 79(04)~\mathrm{km~s^{-1}}$,
and the ephemeris to
\begin{equation}
T_0 = {\rm HJD}~2\,454\,196.7045(25) + 0.362(08)~E,
\label{ec12eph_eq}
\end{equation}
with $E$ giving the cycle number. This timing corresponds to the
inferior conjunction of the emission source, i.e.\ most probably the
secondary star.  A recent radial velocity study by
\citet{maxtedetal07-2} determined $P = 0.3664$ d, which agrees well with
our result.

As explained in Section \ref{genspec_sect}, the non-photometric conditions
cause differences in brightness and continuum slopes between the two nights
(Fig.\ \ref{avsp_fig}, top). Folding the nightly average spectra with 
\citet{bessell90-1} passbands, we find $V = 16.0$ for April 3 and $V = 16.5$ 
for April 5, while $B\!-\!V = 0.2$ for both nights. \citet{kilkennyetal97-1} 
in their discovery paper give $V = 16.20$. However, for EC\,12477--1738 
the emission lines with respect to the continuum also are significantly weaker. 
In Fig.\ \ref{ec12eqw_fig} we see that, while the equivalent widths 
show some variation throughout both nights that does not appear to 
be correlated with the orbital period, the average equivalent width 
drops to almost half its value from April 3 to April 5. The difference 
in the emission line strengths even probably makes up for the difference 
in the continuum slope between the two average spectra (as mentioned
above, it is slightly bluer for April 3), thus yielding identical photometric
colour indices for both nights, while for EC\,13349--3237 and
EC\,14329--1625 the colour indices faithfully reflect the different
continuum slopes. 

The variable line strength suggests that EC\,12477--1738 is another
member of the group of pre-CVs with an active secondary star
\citep[e.g.,][ for \object{EC 13471--1258}, LTT 560, and \object{DE CVn}, 
respectively]{odonoghueetal03-1,tappertetal07-1,vandenbesselaaretal07-1}.
Further long-term photometric or spectroscopic monitoring will be needed for
confirmation.

We have used the spectroscopic decomposition/fit procedure developed
by \citet{rebassa-mansergasetal07-1} to estimate the stellar
parameters of the white dwarf and the companion star in
EC\,12477--1738. In brief, the observed pre-CV spectrum is first
decomposed into a white dwarf and an M-dwarf component using a
$\chi^2$-fit and a set of both white dwarf and M-dwarf template
spectra from the SDSS. After subtracting the best-fit M-dwarf, the
residual spectrum is then subjected to a fit with white dwarf model
spectra computed using the code of \citet{koesteretal05-1}. The model
fit is carried out on the normalised Balmer line profiles to avoid
problems due to uncertainties in the response function of the
spectrograph. Finally, the slope of the continuum is used to break the
degeneracy between "cold" and "hot" model solutions that have
approximatively equally strong Balmer absorption lines. For full
details, we refer the reader to
\citet{rebassa-mansergasetal07-1}. Free parameters in this
decomposition/fit are the white dwarf temperature and surface gravity,
which can be converted to a white dwarf mass using an updated version
of the evolution sequences of \citet{bergeronetal95-2}, the
spectral type of the companion star, as well as distance estimates
based on the flux scaling factors for both stellar components. The
results from the decomposition of both spectra of EC\,12477--1738 are
reported in Table\,\ref{specfit_tab}, and the composite fit to the
April 3 spectrum is shown in Fig.\,\ref{ecspecfit_fig} (top panel).
Despite the different (and non-perfect) atmospheric conditions during
both nights, both fits agree well within the errors, and suggest that
EC\,12477--1738 contains a white dwarf with a temperature of
$T_\mathrm{wd}=17718\pm865$\,K and a mass close to the average mass
of single white dwarfs \citep[e.g.][]{liebertetal05-2},
$M_\mathrm{wd}=0.61\pm0.08\,M_\odot$. The spectral type of the
companion star is M$3\pm0.5$. In principle, the two distance
estimates determined from the flux scaling factor of each component
should agree, but we find $d_\mathrm{sec}>d_\mathrm{wd}$.
\citet{rebassa-mansergasetal07-1} showed that in about $\sim1/3$ of
the systems analysed in their study displayed the same problem. They
discussed possible issues with the adopted spectral type-radius
relation that may be related to stellar activity. Here we note that
EC\,12477--1738 exhibits strong emission line which vary on
time scales of days, consistent with substantial chromospheric
activity on the companion star.

\subsection{EC\,13349--3237}

\begin{figure}
\includegraphics[angle=-90,width=\columnwidth]{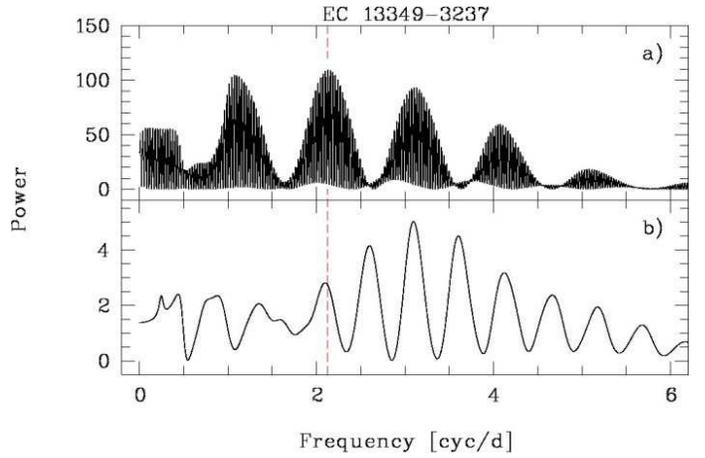}
\caption[]{Scargle periodograms for EC\,13349--3237. a) Photometric data. 
b) Cross-correlation radial velocities 
of the absorption forest in-between $\lambda$5163--5338.
The dashed line indicates the 
adopted period $P = 0.4695~\mathrm{d}$.}
\label{ec13pg_fig}
\end{figure}

\begin{figure}
\includegraphics[angle=-90,width=\columnwidth]{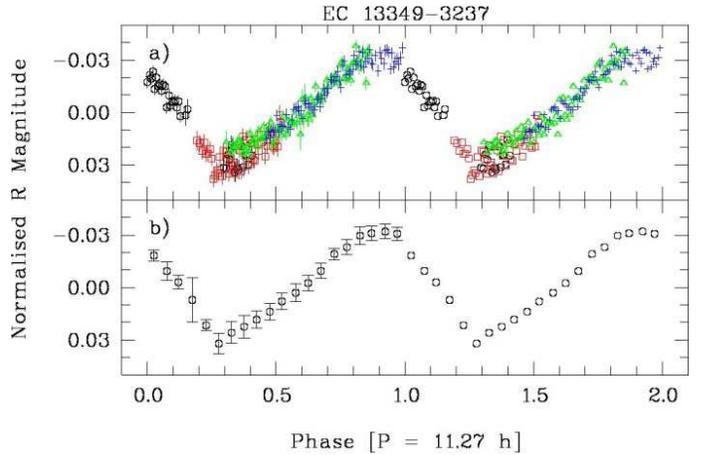}
\caption[]{Phase-folded light curves for EC\,13349--3237. a) The individual data
from May 15 (circles), 16 (squares), June 20 (triangles) and 11 (crosses). 
Two cycles are shown, the second one without error bars. b) The data averaged 
into bins of 0.05 phases.  The error bar gives the standard deviation with respect 
to the average value.}
\label{ec13phlc_fig}
\end{figure}

\begin{figure}
\includegraphics[angle=-90,width=\columnwidth]{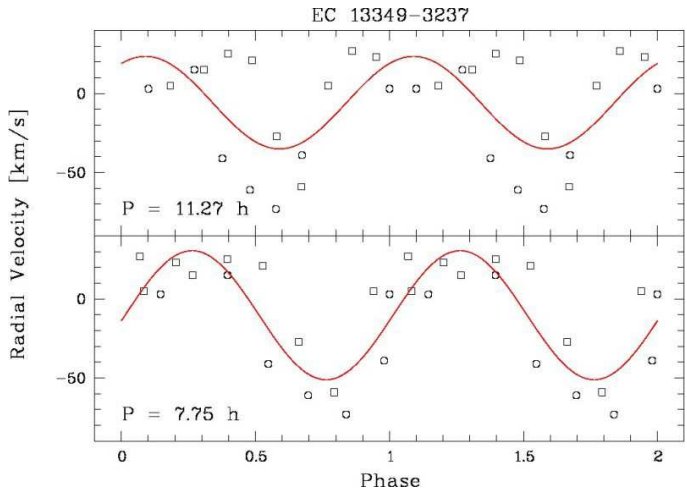}
\caption[]{
EC\,13349--3237 cross-correlation radial velocities folded on the
photometric period (top) and on the "spectroscopic" period (bottom). Symbols 
are the same as in Fig.\ \ref{ec13rvs_fig}. The lines shows the best 
sinusoidal fit with parameters given in Table \ref{rvpar_tab}.}
\label{ec13phrv_fig}
\end{figure}

\begin{figure}
\includegraphics[angle=-90,width=\columnwidth]{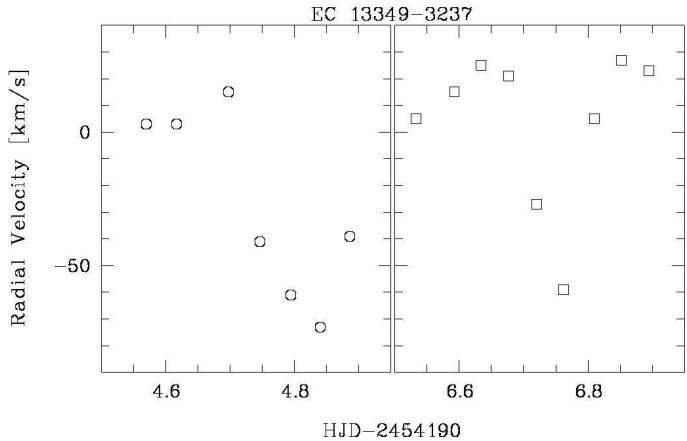}
\caption[]{EC\,13349--3237 radial velocities, as determined by a 
cross-correlation of the absorption forest in-between $\lambda$5163--5338 
{\AA} with a synthetic template.}
\label{ec13rvs_fig}
\end{figure}

Four photometric data sets were taken in two different months, each
time on two subsequent nights. The corresponding periodogram 
(Fig.\ \ref{ec13pg_fig}, top) presents several hubs of fine-spaced 
alias periods. We can discard all hubs longwards of $f = 2.5$, since 
the longest continuous data set of 6.28 h clearly does not represent 
2/3 of an orbit (the triangles in Fig.\ \ref{ec13phlc_fig}). The hub
shortwards of $f = 1.5$ yields an ellipsoidal light curve at an
orbital period $P \sim 22~\mathrm{h}$. At a spectral type of M1 (see below)
this would require an evolved secondary, and there is no spectroscopic
evidence that would support such a scenario. The hub centred at
$f = 2.13$ therefore remains as the only possibility. We have folded
the photometric data with all periods with peak values larger
than half the value of the strongest peak, covering a frequency range
$1.907-2.381~\mathrm{cyc/d}$. Based on the criterion of how the 
data sets of different nights fit together in the phase-folded data,
we find that only two periods, $P_1 = 0.4757~\mathrm{d}$ and $P_2 = 
0.4695~\mathrm{d}$ yield an acceptable light curve. Since $P_2$ is
the slightly stronger one of the two, we adopt as photometric period
$P_\mathrm{ph} = 0.4695(01)~\mathrm{d}$. As a word of caution we
remark that our criterion here assumes that each data set represents
a part of a stable, identical light curve. However, the potential 
presence of star spots or activity on the secondary star could induce
a certain variability of the light curve. This applies to all three
targets of this study, but bears special importance for EC\,13349--3237,
as here we are dealing with 4 incomplete parts of a light curve within two
data sets that are separated by one month.

Somewhat surprisingly, the spectroscopic data do not present a similarly
clear variation, and in fact do not appear to reflect the photometric
variation at all. Measuring radial velocities by fitting single Gaussians to 
the H$\alpha$ emission line or to a number of absorption lines 
(Na\,{\sc I}$\lambda$5893, Ca\,{\sc I}$\lambda$6103 and $\lambda$6122) 
results in very noisy curves without any clear periodic signal. In a second 
attempt we measured radial velocities by cross-correlation in the spectral 
region 5160--5340\,{\AA}, which contains a forest of absorption lines from the 
secondary star due to Mg, Cr and Fe. We used a synthetic template spectrum 
to avoid introducing additional noise into the results. The template was 
calculated using the {\sc uclsyn} code \citep{smalleyetal01-1} and adopting 
$T_\mathrm{eff} = 3500~\mathrm{K}$ and $\log g = 4.5$. This yielded radial 
velocities which were less noisy but still did not demonstrate the expected 
variations in that they do not appear to follow the photometric period, but 
instead prefer $P = 0.323~\mathrm{d}$ (Fig.\ \ref{ec13pg_fig}, bottom). 

We have folded the radial velocity data on both the photometric period
and the one extracted from the spectroscopic periodogram. As expected, since
the photometric period is barely, if at all, present in the spectroscopic
data, that period yields a very poor fit (Fig.\ \ref{ec13phrv_fig}, top). 
The "spectroscopic" period at first glance provides an acceptable fit
to the data (Fig.\ \ref{ec13phrv_fig}, bottom). However, closer inspection 
reveals that there are systematic differences between the data from the 
two nights, as, with one exception, the velocities from the first night 
all lie below the fit. In Fig.\ \ref{ec13rvs_fig} we have plotted the radial
velocities in sequence versus time, which makes it even more obvious that
the velocities do not follow a well-defined sinusoidal variation. We therefore
doubt the physical relevance of this signal. Again we point out that the 
longest photometric data set excludes the "spectroscopic" period for the light 
curve.

Without more 
and better data, we are not able to clarify this puzzling behaviour. Perhaps 
it is due to a combination of the low spectral resolution and a low 
inclination (for the photometric variation the low inclination could be 
compensated for by a particularly strong reflection effect due to a hot white 
dwarf). Further investigation of this system clearly requires time-resolved 
high-resolution spectroscopy.

Folding the nightly average spectra with Bessell filters we obtain
$V = 16.26$ and $B\!-\!V = 0.36$ for April 3, and $V = 16.61$, $B\!-\!V = 0.42$
for April 5. The difference in magnitude is very similar to that found
for EC\,12477--1738, and we attribute this and the difference in the continuum
slope to the non-photometric conditions during the observations. 
Previously reported values for
EC\,13349--3237 are $V = 16.34$, $B\!-\!V = 0.36$ \citep{kilkennyetal97-1}.

Using the spectroscopic decomposition/fit technique introduced in
Sect.\,\ref{ec12477_sect}, we determine the white dwarf temperature
and mass of EC\,13349--3237, $T_\mathrm{wd}=35010\pm3415$\,K, and
$M_\mathrm{wd}=0.46\pm0.11$ (Table\,\ref{specfit_tab} \&
Fig.\,~\ref{ecspecfit_fig}, middle panel). Taken at face value, the
white dwarf mass is lower than the average mass of single white
dwarfs, suggestive of a He-core as a result of the common envelope
evolution. Only a handful of bona-fide He-core white dwarfs in pre-CVs
are known, and a more detailed study appears warranted to confirm this
hypothesis for EC\,13349--3237. The spectral type of the companion,
Sp(sec)\,=\,M1$\pm0.5$. Hence, EC\,13349--3237 is a new addition to
the still very small number of pre-CVs with early-type companion stars
that will start mass transfer above the period gap
\citep{schreiber+gaensicke03-1}. The distances determined for the two
components are in good agreement.

\subsection{EC\,14329--1625}

\begin{figure}
\includegraphics[angle=-90,width=\columnwidth]{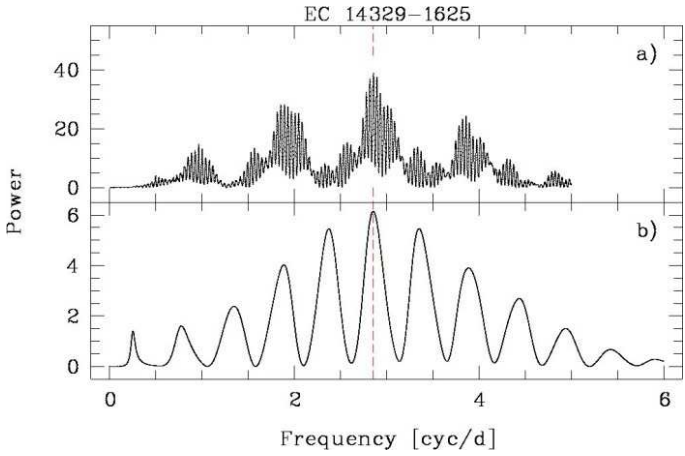}
\caption[]{Scargle periodograms for EC\,14329--1625. a) 2005
photometric data. b) H$\alpha$ radial velocities.
The hashed line marks the adopted period $P = 0.3500~\mathrm{d}$.}
\label{ec14pg_fig}
\end{figure}

\begin{figure}
\includegraphics[angle=-90,width=\columnwidth]{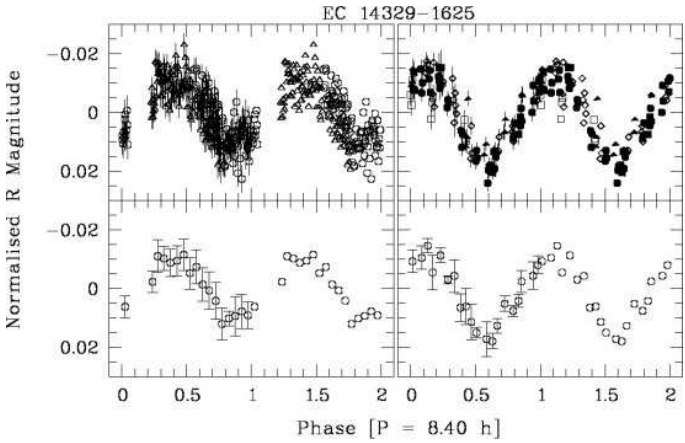}
\caption[]{EC\,14329--1625 phase-folded light curves from 2003 (left) and 2005
(right). Different symbols mark data from different nights. The bottom plots show 
the respective data averaged into 0.05 phase bins.}
\label{ec14phlc_fig}
\end{figure}

\begin{figure}
\includegraphics[angle=-90,width=\columnwidth]{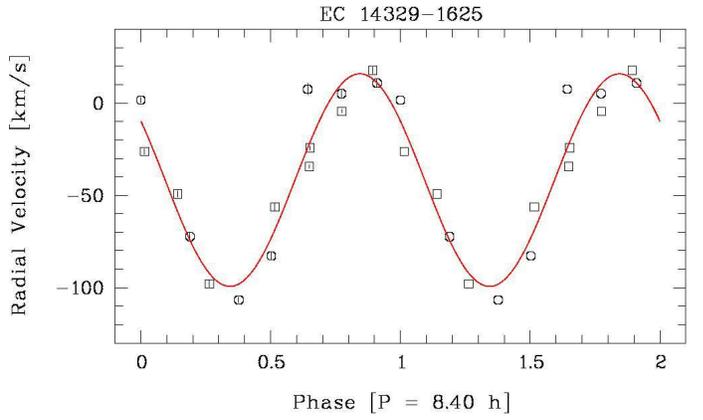}
\caption[]{Phase-folded radial velocities of the H$\alpha$ emission line for 
EC\,14329--1625. Circles mark data from 2007-04-03, squares those from 2007-04-05. 
The solid curve gives the best sine fit. The first cycle includes error bars 
corresponding to the Gaussian fit to the H$\alpha$ emission line.}
\label{ec14phrv_fig}
\end{figure}

This is the brightest of the three pre-CVs of our study, and this made
it possible to also use the photometric data from the 2005 runs which
suffered from bad weather conditions. Unfortunately, the time
span between both sets (from 2003 and 2005) is too long for a combined
period search.  The periodogram of the 2005 data yields as the most
probable period $P_1 = 0.3500(08)~\mathrm{d}$, that agrees well with
the result $P_\mathrm{sp} = 0.3498(35)~\mathrm{d}$ from the radial
velocities of the H$\alpha$ emission line from the 2007 time-resolved
spectroscopy (Fig.\ \ref{ec14pg_fig}). The phase-folded photometric
and spectroscopic data are given in Figs.\ \ref{ec14phlc_fig} and
\ref{ec14phrv_fig}, respectively. The second highest peak for the
photometric data, $P_2 = 0.3457(08)~\mathrm{d}$, yields a visually
equally good light curve, and further observations will be necessary to
definitely break this alias degeneracy. However, since $P_1$ agrees
slightly better with the spectroscopic period, we adopt
$P_\mathrm{orb} = 0.3500(08)~\mathrm{d} = 8.40(02)~\mathrm{h}$ as the
orbital period of EC\,14329--1625. A sine fit to the radial velocity
data yield the parameters listed in Table \ref{rvpar_tab}, and the
corresponding ephemeris is
\begin{equation}
T_0 = {\rm HJD}~2\,454\,196.5437(24) + 0.3500(08)~E,
\end{equation}
with respect to the inferior conjunction of the emission source.

As for the other two targets, the average spectra show a flux difference
between the two nights (Fig,\ \ref{avsp_fig}, bottom). Folding the data 
with Bessell filter curves
yields similar differences as for the other systems, with
values $V = 14.79$, $B\!-\!V = 0.33$ for April 3, and $V = 15.13$, 
$B\!-\!V = 0.38$ for April 5. \citet{kilkennyetal97-1} report
$V = 14.89$, $B\!-\!V = 0.25$, and again the April 3 data agree slightly
better with their measurements.

For EC\,14329--1625, the spectroscopic decomposition (see
Sect.\,\ref{ec12477_sect}) results in $T_\mathrm{wd}=14600\pm1300$\,K
and $M_\mathrm{wd}=0.62\pm0.14$ (Table\,\ref{specfit_tab} \&
Fig.\,~\ref{ecspecfit_fig}, middle panel), and similar to
EC\,12477--1738, EC\,14329--1625 has a mass close to the average mass
of single white dwarfs.  Another similarity to EC\,12477--1738 is that
we find $d_\mathrm{sec}>d_\mathrm{wd}$, which may suggest that the
companion star has a radius slightly too large for its spectral type.
As already mentioned in Sect.\,\ref{ec12477_sect}, this might be
related to stellar activity on the companion star
\citep{rebassa-mansergasetal07-1}. In fact, both EC\,12477--1738 and
EC\,14329--1625 exhibit very strong Balmer emission lines. Given their
long orbital periods and modest white dwarf temperatures, the strength
of the Balmer lines is indicative of chromospheric activity, rather
than irradiation/heating of the companion star. 

\section{Conclusions}

\begin{table}
\caption[]{Radial-velocity parameters: the adopted orbital period $P$, the
semi-amplitude $K_2$, the constant term $\gamma$, and the standard deviation of
the fit $\sigma$. 
}
\label{rvpar_tab}
\begin{tabular}{lllll}
\hline\noalign{\smallskip}
object & $P$ & $K_2$ & $\gamma$ & $\sigma$ \\
       & [d] & [km s$^{-1}$] & [km s$^{-1}$] & [km s$^{-1}$] \\
\hline\noalign{\smallskip}
EC\,12477--1738 & 0.362(08)  & 79(04) & 29(03)    & 19 \\
EC\,13349--3237 & 0.4695(01) & [29(06) & $-$6(04)  & 27] \\
                & 0.4757(01)$^1$ &      &           &    \\
                & [0.323(08) & 41(04) & $-$10(03) & 20]$^2$ \\
EC\,14329--1625 & 0.3500(08) & 58(03) & $-$42(02) & 12 \\
                & 0.3457(08)$^1$ &        &           &    \\
\hline\noalign{\smallskip}
\multicolumn{5}{l}{\parbox{8.3cm}{
1) best alias period, yielding an equally acceptable light curve;
2) the "spectroscopic" period and the radial 
velocity parameters are formally included, but should be taken with great 
caution, due to the radial-velocity variation being not well understood
}}
\end{tabular}
\end{table}

We have determined the system parameters for the three pre-CVs EC\,12477--1738,
EC\,13349--3237, and EC\,14329--1625. The results of the spectroscopic 
decomposition
are given in Table \ref{specfit_tab}, the parameters gained from the radial
velocities are summarised in Table \ref{rvpar_tab}. When comparing our results
with previously published data we find that \citet{koesteretal01-1} determine
somewhat higher temperatures for the WDs in EC\,12477--1738 
($20\,922\pm317$\,K vs $17\,718\pm865$\,K) and in EC\,13349--3237
($48\,116\pm1\,353$\,K vs  $35\,010\pm3\,000$\,K). These differences
are explained by the fact that \citet{koesteretal01-1} did
not correct for the contribution of the companion star before fitting
the white dwarf spectrum. Consequently, the equivalent widths of
the white dwarf photospheric Balmer lines are underestimated, which
pulls the spectroscopic fit to higher temperatures (a higher degree of
ionisation). 

The secondary spectral types for all three pre-CVs agree well with those obtained
by \citet{tappertetal07-2} from absorption line strengths in $K$-band
spectra: M3V (this study: M3$\pm$0.5V) for EC\,12477--1738 and 
EC\,14329--1625, and K5V--M2V (M1$\pm$0.5V) for EC\,13349--3237.

Two of our systems, EC\,12477--1738 and EC\,14329--1625, turn out to have 
very similar parameters: their orbital period is close to 8.5 h, and their
stellar components are a medium-hot WD and an M3V secondary. The latter is
an unusual combination for the pre-SDSS sample of pre-CVs, which was strongly
biased to hotter WDs \citep{schreiber+gaensicke03-1}, but it is now found quite
frequently in pre-CVs discovered in the SDSS \citep{rebassa-mansergasetal07-1}.
EC\,13349--3237, on the other hand, while submitting to the usual observational
bias in being a young pre-CV with a hot WD, also incorporates a comparatively 
early-type, M1V, secondary star. Pre-CVs with an early-type secondary star
still represent a minority in the currently known sample. As
\citet{schreiber+gaensicke03-1} point out, this is due to the fact that
the secondary star in these systems contributes too much light for the object
to appear as a candidate WD or QSO in the colour-colour diagram and thus will
not trigger spectroscopic follow-up observations. It is probable that, if
EC\,13349--3237 contained a cooler WD, it would still remain undiscovered 
\citep[see also][]{schreiberetal07-1}.

Based on the spectral type of the secondary stars, we can expect that EC\,12477--1738
and EC\,14329--1625 will turn into CVs with an orbital period 3--5 h, while
EC\,13349--3237 will enter its CV phase at $P \sim 4-6~{\mathrm h}$ 
\citep[e.g.,][]{beuermannetal98-1}. In this context it is worth noticing that
the period regime at 3--4 h is dominated by the high-mass-transfer SW Sex stars,
whose physics and possible magnetic nature are still under debate 
\citep[][ and references herein]{rodriguez-giletal07-1}. A detailed
examination of pre-CVs with secondary spectral types $\sim$M3V, such as
EC\,12477--1738 and EC\,14329--1625, might therefore provide valuable insight
into this important subgroup of CVs.

\begin{acknowledgements}
We thank the anonymous referee for comments that helped to improve the paper.
CT and REM acknowledge financial support by FONDECYT grant 1051078. REM also
acknowledges financial support by the Chilean Center for Astrophysics 
FONDAP 15010003 and  from the BASAL Centro de Astrof\'{\i}sica y Tecnologias 
Afines (CATA) PFB--06/2007. We thank Detlev Koester for providing us his 
white dwarf model spectra.  This work has made intensive use of the SIMBAD 
database, operated at CDS, Strasbourg, France, and of NASA's Astrophysics 
Data System Bibliographic Services. The Digitized Sky Surveys were produced 
at the Space Telescope Science Institute under U.S. Government grant 
NAG W-2166, based on photographic data obtained using the Oschin Schmidt 
Telescope on Palomar Mountain and the UK Schmidt Telescope. IRAF is 
distributed by the National Optical Astronomy Observatories.
\end{acknowledgements}


\begin{thebibliography}{40}
\expandafter\ifx\csname natexlab\endcsname\relax\def\natexlab#1{#1}\fi

\bibitem[{{Bergeron} {et~al.}(1995){Bergeron}, {Wesemael}, \&
  {Beauchamp}}]{bergeronetal95-2}
{Bergeron}, P., {Wesemael}, F., \& {Beauchamp}, A. 1995, \pasp, 107, 1047

\bibitem[{{Bessell}(1990)}]{bessell90-1}
{Bessell}, M.~S. 1990, \pasp, 102, 1181

\bibitem[{{Beuermann} {et~al.}(1998){Beuermann}, {Baraffe}, {Kolb}, \&
  {Weichhold}}]{beuermannetal98-1}
{Beuermann}, K., {Baraffe}, I., {Kolb}, U., \& {Weichhold}, M. 1998, \aap, 339,
  518

\bibitem[{{Broeg} {et~al.}(2005){Broeg}, {Fern{\'a}ndez}, \&
  {Neuh{\"a}user}}]{broegetal05-1}
{Broeg}, C., {Fern{\'a}ndez}, M., \& {Neuh{\"a}user}, R. 2005, Astronomische
  Nachrichten, 326, 134

\bibitem[{{G{\"a}nsicke}(2005)}]{gaensicke05-1}
{G{\"a}nsicke}, B.~T. 2005, in Astronomical Society of the Pacific Conference
  Series, Vol. 330, The Astrophysics of Cataclysmic Variables and Related
  Objects, ed. J.-M. {Hameury} \& J.-P. {Lasota}

\bibitem[{{G{\"a}nsicke} {et~al.}(2003){G{\"a}nsicke}, {Szkody}, {de Martino},
  {Beuermann}, {Long}, {Sion}, {Knigge}, {Marsh}, \&
  {Hubeny}}]{gaensickeetal03-1}
{G{\"a}nsicke}, B.~T., {Szkody}, P., {de Martino}, D., {et~al.} 2003, \apj,
  594, 443

\bibitem[{{Harrison} {et~al.}(2004){Harrison}, {Osborne}, \&
  {Howell}}]{harrisonetal04-1}
{Harrison}, T.~E., {Osborne}, H.~L., \& {Howell}, S.~B. 2004, \aj, 127, 3493

\bibitem[{{Harrison} {et~al.}(2005){Harrison}, {Osborne}, \&
  {Howell}}]{harrisonetal05-1}
{Harrison}, T.~E., {Osborne}, H.~L., \& {Howell}, S.~B. 2005, \aj, 129, 2400

\bibitem[{{Horne}(1986)}]{horne86-1}
{Horne}, K. 1986, \pasp, 98, 609

\bibitem[{{Kilkenny} {et~al.}(1997){Kilkenny}, {O'Donoghue}, {Koen}, {Stobie},
  \& {Chen}}]{kilkennyetal97-1}
{Kilkenny}, D., {O'Donoghue}, D., {Koen}, C., {Stobie}, R.~S., \& {Chen}, A.
  1997, \mnras, 287, 867

\bibitem[{{Koester} {et~al.}(2001){Koester}, {Napiwotzki}, {Christlieb},
  {Drechsel}, {Hagen}, {Heber}, {Homeier}, {Karl}, {Leibundgut}, {Moehler},
  {Nelemans}, {Pauli}, {Reimers}, {Renzini}, \& {Yungelson}}]{koesteretal01-1}
{Koester}, D., {Napiwotzki}, R., {Christlieb}, N., {et~al.} 2001, \aap, 378,
  556

\bibitem[{{Koester} {et~al.}(2005){Koester}, {Napiwotzki}, {Voss}, {Homeier},
  \& {Reimers}}]{koesteretal05-1}
{Koester}, D., {Napiwotzki}, R., {Voss}, B., {Homeier}, D., \& {Reimers}, D.
  2005, \aap, 439, 317

\bibitem[{{Larsson}(1996)}]{larsson96-1}
{Larsson}, S. 1996, \aaps, 117, 197

\bibitem[{{Liebert} {et~al.}(2005){Liebert}, {Bergeron}, \&
  {Holberg}}]{liebertetal05-2}
{Liebert}, J., {Bergeron}, P., \& {Holberg}, J.~B. 2005, \apjs, 156, 47

\bibitem[{{Maxted} {et~al.}(2007){Maxted}, {Napiwotzki}, {Marsh}, {Burleigh},
  {Dobbie}, {Hogan}, \& {Nelemans}}]{maxtedetal07-2}
{Maxted}, P.~F.~L., {Napiwotzki}, R., {Marsh}, T.~R., {et~al.} 2007, in
  Astronomical Society of the Pacific Conference Series, Vol. 372, 15th
  European Workshop on White Dwarfs, ed. R.~{Napiwotzki} \& M.~R. {Burleigh},
  471

\bibitem[{{Napiwotzki} \& {Burleigh}(2007)}]{napiwotzki+burleigh07-1}
{Napiwotzki}, R. \& {Burleigh}, M.~R., eds. 2007, Astronomical Society of the
  Pacific Conference Series, Vol. 372, {15th European Workshop on White Dwarfs}

\bibitem[{{O'Donoghue} {et~al.}(2003){O'Donoghue}, {Koen}, {Kilkenny},
  {Stobie}, {Koester}, {Bessell}, {Hambly}, \&
  {MacGillivray}}]{odonoghueetal03-1}
{O'Donoghue}, D., {Koen}, C., {Kilkenny}, D., {et~al.} 2003, \mnras, 345, 506

\bibitem[{{Patterson}(1998)}]{patterson98-1}
{Patterson}, J. 1998, \pasp, 110, 1132

\bibitem[{{Raymond} {et~al.}(2003){Raymond}, {Szkody}, {Hawley}, {Anderson},
  {Brinkmann}, {Covey}, {McGehee}, {Schneider}, {West}, \&
  {York}}]{raymondetal03-1}
{Raymond}, S.~N., {Szkody}, P., {Hawley}, S.~L., {et~al.} 2003, \aj, 125, 2621

\bibitem[{{Rebassa-Mansergas} {et~al.}(2007){Rebassa-Mansergas},
  {G{\"a}nsicke}, {Rodr{\'{\i}}guez-Gil}, {Schreiber}, \&
  {Koester}}]{rebassa-mansergasetal07-1}
{Rebassa-Mansergas}, A., {G{\"a}nsicke}, B.~T., {Rodr{\'{\i}}guez-Gil}, P.,
  {Schreiber}, M.~R., \& {Koester}, D. 2007, \mnras, 382, 1377

\bibitem[{{Ritter} \& {Kolb}(2003)}]{ritter+kolb03-1}
{Ritter}, H. \& {Kolb}, U. 2003, \aap, 404, 301

\bibitem[{{Rodr{\'{\i}}guez-Gil} {et~al.}(2007){Rodr{\'{\i}}guez-Gil},
  {G{\"a}nsicke}, {Hagen}, {Araujo-Betancor}, {Aungwerojwit}, {Allende Prieto},
  {Boyd}, {Casares}, {Engels}, {Giannakis}, {Harlaftis}, {Kube}, {Lehto},
  {Mart{\'{\i}}nez-Pais}, {Schwarz}, {Skidmore}, {Staude}, \&
  {Torres}}]{rodriguez-giletal07-1}
{Rodr{\'{\i}}guez-Gil}, P., {G{\"a}nsicke}, B.~T., {Hagen}, H.-J., {et~al.}
  2007, \mnras, 377, 1747

\bibitem[{{Scargle}(1982)}]{scargle82-1}
{Scargle}, J.~D. 1982, \apj, 263, 835

\bibitem[{{Schenker} \& {King}(2002)}]{schenker+king02-1}
{Schenker}, K. \& {King}, A.~R. 2002, in Astronomical Society of the Pacific
  Conference Series, Vol. 261, The Physics of Cataclysmic Variables and Related
  Objects, ed. B.~T. {G{\"a}nsicke}, K.~{Beuermann}, \& K.~{Reinsch}, 242

\bibitem[{{Schenker} {et~al.}(2002){Schenker}, {King}, {Kolb}, {Wynn}, \&
  {Zhang}}]{schenkeretal02-1}
{Schenker}, K., {King}, A.~R., {Kolb}, U., {Wynn}, G.~A., \& {Zhang}, Z. 2002,
  \mnras, 337, 1105

\bibitem[{{Schreiber} {et~al.}(2007){Schreiber}, {Nebot Gomez-Moran}, \&
  {Schwope}}]{schreiberetal07-1}
{Schreiber}, M., {Nebot Gomez-Moran}, A., \& {Schwope}, A. 2007, in
  Astronomical Society of the Pacific Conference Series, Vol. 372, 15th
  European Workshop on White Dwarfs, ed. R.~{Napiwotzki} \& M.~R. {Burleigh},
  459

\bibitem[{{Schreiber} \& {G{\" a}nsicke}(2003)}]{schreiber+gaensicke03-1}
{Schreiber}, M.~R. \& {G{\" a}nsicke}, B.~T. 2003, \aap, 406, 305

\bibitem[{{Schreiber} {et~al.}(2008){Schreiber}, {G{\"a}nsicke}, {Southworth},
  {Schwope}, \& {Koester}}]{schreiberetal08-1}
{Schreiber}, M.~R., {G{\"a}nsicke}, B.~T., {Southworth}, J., {Schwope}, A.~D.,
  \& {Koester}, D. 2008, \aap, 484, 441

\bibitem[{{Schwarzenberg-Czerny}(1989)}]{schwarzenberg-czerny89-1}
{Schwarzenberg-Czerny}, A. 1989, \mnras, 241, 153

\bibitem[{{Schwarzenberg-Czerny}(1996)}]{schwarzenberg-czerny96-1}
{Schwarzenberg-Czerny}, A. 1996, \apjl, 460, L107

\bibitem[{{Silvestri} {et~al.}(2006){Silvestri}, {Hawley}, {West}, {Szkody},
  {Bochanski}, {Eisenstein}, {McGehee}, {Schmidt}, {Smith}, {Wolfe}, {Harris},
  {Kleinman}, {Liebert}, {Nitta}, {Barentine}, {Brewington}, {Brinkmann},
  {Harvanek}, {Krzesi{\'n}ski}, {Long}, {Neilsen}, {Schneider}, \&
  {Snedden}}]{silvestrietal06-1}
{Silvestri}, N.~M., {Hawley}, S.~L., {West}, A.~A., {et~al.} 2006, \aj, 131,
  1674

\bibitem[{{Silvestri} {et~al.}(2007){Silvestri}, {Lemagie}, {Hawley}, {West},
  {Schmidt}, {Liebert}, {Szkody}, {Mannikko}, {Wolfe}, {Barentine},
  {Brewington}, {Harvanek}, {Krzesinski}, {Long}, {Schneider}, \&
  {Snedden}}]{silvestrietal07-1}
{Silvestri}, N.~M., {Lemagie}, M.~P., {Hawley}, S.~L., {et~al.} 2007, \aj, 134,
  741

\bibitem[{{Smalley} {et~al.}(2001){Smalley}, {Smith}, \&
  {Dworetsky}}]{smalleyetal01-1}
{Smalley}, B., {Smith}, K., \& {Dworetsky}, M. 2001, {\sc UCLSYN} Userguide

\bibitem[{{Stehle} {et~al.}(1997){Stehle}, {Kolb}, \&
  {Ritter}}]{stehleetal97-1}
{Stehle}, R., {Kolb}, U., \& {Ritter}, H. 1997, \aap, 320, 136

\bibitem[{{Stetson}(1992)}]{stetson92-1}
{Stetson}, P.~B. 1992, in Astronomical Society of the Pacific Conference
  Series, Vol.~25, Astronomical Data Analysis Software and Systems I, ed. D.~M.
  {Worrall}, C.~{Biemesderfer}, \& J.~{Barnes}, 297

\bibitem[{{Tappert} {et~al.}(2004){Tappert}, {G{\"a}nsicke}, \&
  {Mennickent}}]{tappertetal04-1}
{Tappert}, C., {G{\"a}nsicke}, B.~T., \& {Mennickent}, R.~E. 2004, in Revista
  Mexicana de Astronomia y Astrofisica Conference Series, Vol.~20, Compact
  Binaries in the Galaxy and Beyond, ed. G.~{Tovmassian} \& E.~{Sion}, 245

\bibitem[{{Tappert} {et~al.}(2007{\natexlab{a}}){Tappert}, {G{\"a}nsicke},
  {Schmidtobreick}, {Aungwerojwit}, {Mennickent}, \&
  {Koester}}]{tappertetal07-1}
{Tappert}, C., {G{\"a}nsicke}, B.~T., {Schmidtobreick}, L., {et~al.}
  2007{\natexlab{a}}, \aap, 474, 205

\bibitem[{{Tappert} {et~al.}(2007{\natexlab{b}}){Tappert}, {G{\"a}nsicke},
  {Schmidtobreick}, {Mennickent}, \& {Navarrete}}]{tappertetal07-2}
{Tappert}, C., {G{\"a}nsicke}, B.~T., {Schmidtobreick}, L., {Mennickent},
  R.~E., \& {Navarrete}, F.~P. 2007{\natexlab{b}}, \aap, 475, 575

\bibitem[{{Tappert} {et~al.}(2006){Tappert}, {Toledo}, {G{\"a}nsicke}, \&
  {Mennickent}}]{tappertetal06-2}
{Tappert}, C., {Toledo}, I., {G{\"a}nsicke}, B.~T., \& {Mennickent}, R.~E.
  2006, in Revista Mexicana de Astronomia y Astrofisica Conference Series,
  Vol.~26, XI IAU Regional Latin American Meeting of Astronomy, ed.
  L.~{Infante} \& M.~{Rubio}, 177

\bibitem[{{van den Besselaar} {et~al.}(2007){van den Besselaar}, {Greimel},
  {Morales-Rueda}, {Nelemans}, {Thorstensen}, {Marsh}, {Dhillon}, {Robb},
  {Balam}, {Guenther}, {Kemp}, {Augusteijn}, \&
  {Groot}}]{vandenbesselaaretal07-1}
{van den Besselaar}, E.~J.~M., {Greimel}, R., {Morales-Rueda}, L., {et~al.}
  2007, \aap, 466, 1031

\end{thebibliography}

\end{document}